\documentclass[aip,rsi,reprint,graphicx]{revtex4-1} 

\usepackage{graphicx}
\usepackage{amsmath}
\usepackage{braket}
\draft 

\begin{document}

\title{A Low Temperature Nonlinear Optical Rotational Anisotropy Spectrometer for the Determination of Crystallographic and Electronic Symmetries}

\author{Darius H. Torchinsky}
\affiliation{Institute for Quantum Information and Matter, California Institute of Technology, Pasadena, CA 91125, USA}
\affiliation{Department of Physics, California Institute of Technology, Pasadena, CA 91125, USA}
\author{Hao Chu}
\affiliation{Institute for Quantum Information and Matter, California Institute of Technology, Pasadena, CA 91125, USA}
\affiliation{Department of Applied Physics, California Institute of Technology, Pasadena, CA 91125, USA}
\author{Tongfei Qi}
\affiliation{Center for Advanced Materials, Department of Physics and Astronomy, University of Kentucky, Lexington, Kentucky 40506, USA}
\author{Gang Cao}
\affiliation{Center for Advanced Materials, Department of Physics and Astronomy, University of Kentucky, Lexington, Kentucky 40506, USA}
\author{David Hsieh}
\affiliation{Institute for Quantum Information and Matter, California Institute of Technology, Pasadena, CA 91125, USA}
\affiliation{Department of Physics, California Institute of Technology, Pasadena, CA 91125, USA}

\date{\today}

\begin{abstract}
Nonlinear optical generation from a crystalline material can reveal the symmetries of both its lattice structure and underlying ordered electronic phases and can therefore be exploited as a complementary technique to diffraction based scattering probes. Although this technique has been successfully used to study the lattice and magnetic structures of systems such as semiconductor surfaces, multiferroic crystals, magnetic thin films and multilayers, challenging technical requirements have prevented its application to the plethora of complex electronic phases found in strongly correlated electron systems. These requirements include an ability to probe small bulk single crystals at the micron length scale, a need for sensitivity to the entire nonlinear optical susceptibility tensor, oblique light incidence reflection geometry and incident light frequency tunability among others. These measurements are further complicated by the need for extreme sample environments such as ultra low temperatures, high magnetic fields or high pressures. In this review we present a novel experimental construction using a rotating light scattering plane that meets all the aforementioned requirements. We demonstrate the efficacy of our scheme by making symmetry measurements on a micron scale facet of a small bulk single crystal of Sr$_2$IrO$_4$ using optical second and third harmonic generation.
\end{abstract}


\maketitle
\section{Introduction}

Determining the symmetry of a crystalline solid and its underlying ordered electronic phases is essential for understanding its macroscopic mechanical, electrical and magnetic properties \cite{Nye, Birss}. X-ray \cite{Warren}, neutron \cite{Squires} and electron diffraction \cite{Zou} have powerful complementary abilities to probe lattice, magnetic and charge symmetries, while resonant x-ray diffraction has demonstrated sensitivity to even more exotic types of symmetry involving ordered orbital~\cite{Murakami} and higher multipolar degrees of freedom~\cite{Santini, Kiss}. However, an accurate symmetry assignment, which relies on being able to perform a unique fit to a diffraction pattern, is not always possible. Technical obstacles include not having a sufficient number of Bragg peaks owing to a finite instrument momentum range; spurious peaks arising from multiple scattering events, parasitic phases or microscopic domains in a crystal; the presence of elements with strong absorption or weak scattering cross-sections; and the unavailability of large single crystals comparable with the probe beam size.

Nonlinear optical generation~\cite{Shen, Boyd} is an alternative non-diffraction based technique for determining the symmetries of the lattice and ordered electronic (electric or magnetic) phases of a crystal. This approach is based on Neumann's principle, which dictates that a tensor representing any physical property of a crystal must be invariant under every symmetry operation of its lattice or underlying electronic order~\cite{Nye, Birss}. These conditions of invariance establish a set of relationships between tensor components that reduce the number that are non-zero and independent. The structure of a tensor response therefore embeds
the symmetries of a crystal, with higher rank tensors allowing for more accurate levels of refinement. Nonlinear optical susceptibility tensors are particularly useful because they are sensitive to both lattice \cite{Tom1,Yang} and electronic symmetries~\cite{Pan,Dahn,Fiebig_Review,Kirilyuk_Review} and because tensors of arbitrary rank can be probed through successively higher nonlinear harmonic generation (NHG) processes in a crystal. Moreover it offers several unique capabilities compared with diffraction based probes including micron scale spatial resolution and bulk versus surface selectivity \cite{Shen_review,Sipe}.

A nonlinear harmonic generation rotational anisotropy (NHG-RA) measurement is typically carried out to determine the structure of a nonlinear optical susceptibility tensor, which involves recording the intensity of high harmonic light generated from a crystal as it rotates about some crystalline axis. However, several technical challenges associated with maintaining precise optical alignment from a rotating sample have so far precluded such experiments from being performed on small bulk single
crystals and under extreme sample environments such as ultra low temperature, high magnetic field or externally imposed strain. In this review, we describe the design, construction and operation of a novel NHG-RA spectrometer that overcomes all these challenges through a rotation of the scattering plane as opposed to the sample. Our setup opens the way to apply NHG-RA to a broad range of materials, including many \textit{d}- and \textit{f}-electron based strongly correlated electron systems, which are typically only available in small bulk single crystalline form. Moreover it allows measurements to be performed in ultra low temperature optical cryostats and under static magnetic or strain fields.

The review is organized as follows. In Section~\ref{sec:bkgrnd} we introduce the theoretical background to NHG responses and their relationship to the structural and electronic symmetries of a crystal. In Section~\ref{sec:others} we describe the capabilities and technical limitations of existing NHG-RA setups. In Section~\ref{sec:design} we describe the design and construction of our NHG-RA spectrometer and present representative results in Section~\ref{sec:results} on a 5$d$ transition metal oxide Sr$_2$IrO$_4$. Finally in Section~\ref{sec:summary} we discuss how the technique can be generalized to an imaging modality to understand crystallographic and electronic domains and how it can be utilized for time-resolved pump-probe experiments.

\section{Nonlinear harmonic generation in crystals}\label{sec:bkgrnd}

Nonlinear harmonic generation is a process by which monochromatic light of frequency $\omega$ is converted into higher harmonics $n\omega$ ($n$ = 2,3,4...) through its nonlinear interaction with a material \cite{Shen, Boyd}. In general, the oscillating electric $\vec{E}(\omega)$ and magnetic $\vec{H}(\omega)$ fields of incident light can induce oscillating electric dipole $\vec{P}(n\omega)$, magnetic dipole $\vec{M}(n\omega)$, electric quadrupole $\tensor{Q}(n\omega)$ or even higher order multipole
densities in a material that act as sources of higher harmonic radiation. Each NHG process is governed by a specific nonlinear optical susceptibility tensor $\tensor{\chi}$ of the material. For example, magnetic dipole second harmonic generation induced via one interaction with both the incident electric and magnetic
fields would conventionally \cite{Fiebig_Review} be expressed as $M_i(2\omega) = \chi^{mem}_{ijk}E_j(\omega)H_k(\omega)$, where the first superscript denotes the magnetic dipole ($m$) origin of the induced source, the second and third superscripts denote the electric ($e$) and magnetic ($m$) nature of the driving fields and the subscripts denote the polarization components.

Microscopically the nonlinear optical susceptibility tensor is expressed via terms such as
\begin{equation}\label{eq:microchi2}
\chi^{mem}_{ijk} \propto \sum_{g,n,n'}
\left[\frac{\langle g |M_i|n\rangle \langle n |P_j|n'\rangle \langle n'|M_k|g\rangle}{(2\omega-\omega_{ng})(\omega-\omega_{n'g})}+\cdots\right]f_g
\end{equation}
which describes a two-photon absorption process driven by a magnetic dipole transition from the initial $|g\rangle$ to intermediate state $|n'\rangle$ and an electric dipole transition from the intermediate $|n'\rangle$ to final state $|n\rangle$, followed by a frequency doubled one-photon emission process driven by a magnetic dipole transition from $|n\rangle$ back to $|g\rangle$~\cite{Fiebig_Review}. The energy difference between the initial and intermediate or final states is given by $\omega_{n'g}$ or $\omega_{ng}$ respectively and $f_g$ is the Fermi distribution function for state $|g\rangle$.

Neumann's principle is applied to $\tensor{\chi}$ by enforcing invariance under transformations that respect both the lattice and electronic symmetries of the crystal, which reduces the number of independent non-zero tensor components. Further reductions can be made for experiments using a single incident beam by exploiting
the permutation symmetry of the incident fields. The lattice and electronic symmetries can therefore in principle be resolved by measuring all components of $\tensor{\chi}$ using frequencies tuned both to and away from optical transitions involving states undergoing electronic ordering.

\section{Conventional NHG-RA system design}\label{sec:others}

\begin{figure}
\includegraphics[scale=0.6]{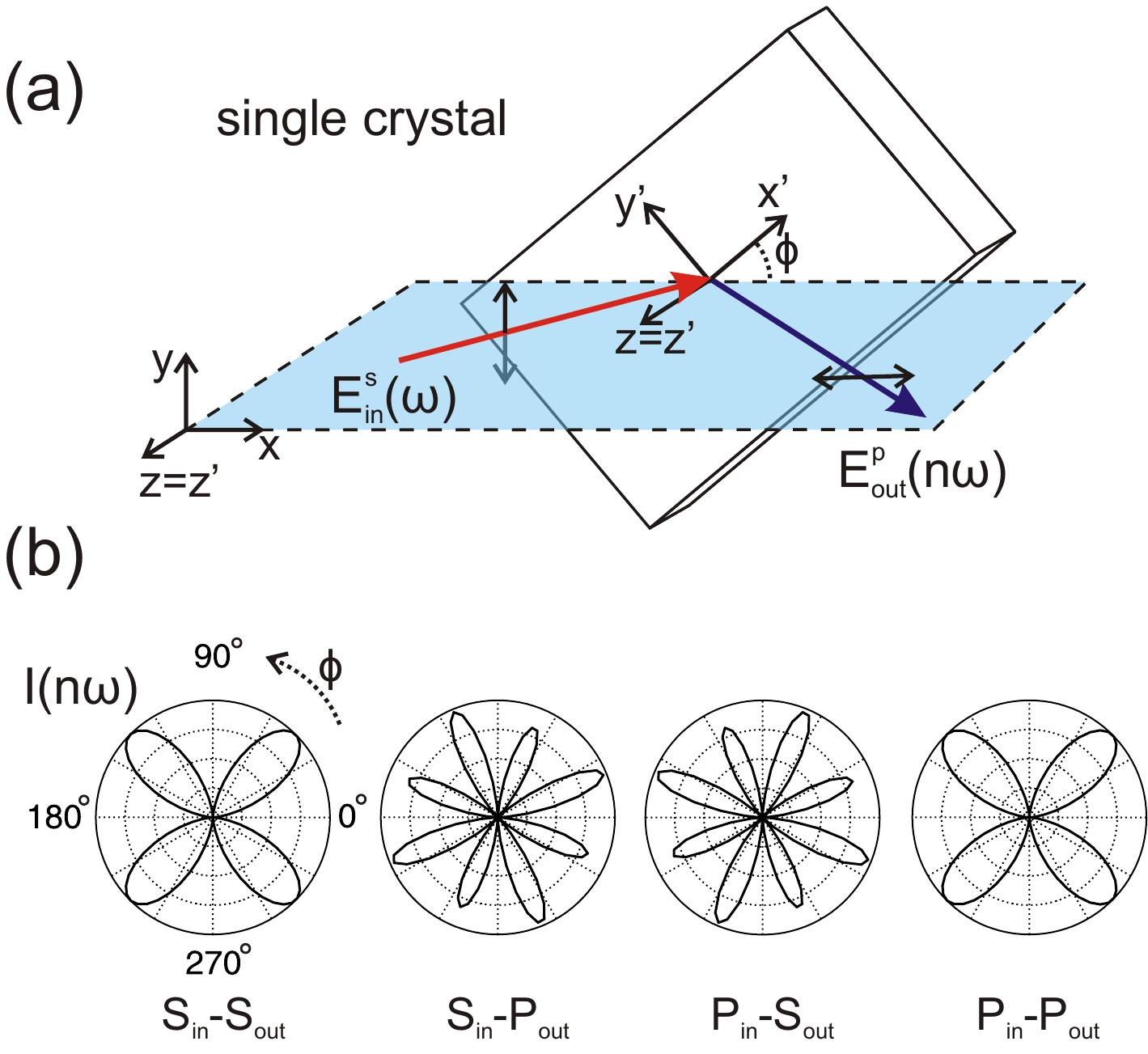}
\caption{\label{fig:splane} (a) Schematic of an NHG-RA experiment. The scattering plane (light blue) is defined by the incident (red arrow) and radiated (dark blue arrow) beams. S(P)-polarization denotes an electric field pointing perpendicular (parallel) to the scattering plane. $\phi$ is the angle that the crystalline axis
$x'$ makes with respect to the scattering plane axis $x$ upon rotation about surface normal $z=z'$ axis. (b) Simulated RA patterns for a $\chi^{qee}_{ijkl}$ process using arbitrary magnitudes for the independent non-zero tensor components allowed from a $D_{4h}$ crystallographic point group.}
\end{figure}

In practice, the components of $\tensor{\chi}$ are measured using a NHG-RA technique where the intensity of high harmonic radiation emitted from a crystal is measured as a function of the angle $\phi$ subtended between the light scattering plane and the crystalline axes (Fig.~\ref{fig:splane}a). Rotational anisotropy patterns measured using different combinations of incident and radiated light polarization (Fig.~\ref{fig:splane}b) and different crystal faces probe distinct combinations of tensor components. Therefore, a collection of RA patterns is typically required to completely determine the structure of $\tensor{\chi}$. Depending on whether the incident and radiated frequencies are tuned off or on resonance with electronic transitions involving states participating in the electronic order, the tensor structure will be primarily representative of the lattice or electronic symmetries, respectively.

To date, NHG-RA experiments have largely been conducted using one of two schemes. In the first scheme, light is normally incident on a crystal face and the transmitted high harmonic radiation is measured. The advantage of this geometry is that RA patterns can be obtained by simply rotating the selected polarizations of the incident and radiated beams, while keeping the crystal stationary. This scheme has proven particularly conducive to studying the symmetry of magnetic \cite{Fiebig_Cr2O3,Gridnev,Fiebig_NiO_PRL,Lafrentz} and multiferroic order \cite{Orenstein,Fiebig_manganite} in thin transparent crystals. It has also been applied to study the lattice structure of opaque crystals by measuring the retro-reflected high harmonic radiation \cite{Heinz,Tom,Petersen,Hirata}. However a limitation is that no incident field component can be introduced perpendicular to the crystal surface, which greatly reduces the number of accessible tensor components.

The second scheme utilizes an oblique reflection geometry where the polarizations of the incident and reflected beams are held fixed while the crystal is rotated (Fig.~\ref{fig:splane}a) to collect a RA pattern. However this requires aligning an optically flat region of the crystal to coincide and lie normal to a manipulator rotation axis, which in turn must be made to lie in the light scattering plane. Owing to the limited number and precision of mechanical degrees of freedom on a typical cryostat manipulator, this scheme is only applicable to thin films \cite{Banshchikov,Sato,Shelford,Kim,An} or bulk single crystals that have several mm large naturally cleaving \cite{Hsieh_SHG} or mechanically polished \cite{Nyvlt} flat areas. Experiments can be simplified by rotating the polarizations of the incident and radiated light while keeping the crystal stationary \cite{Sheu,Ogawa}, although this can restrict the number of accessible tensor components because the scattering plane stays fixed with respect to the crystalline axes.

However a challenging combination of technical requirements have so far prevented NHG-RA from being widely applied to the study of complex low temperature electronic phases. These include the following: i) Experiments must be performed in reflection geometry because the thickness of bulk single crystals typically exceeds the penetration depth of light especially at inter-band resonance frequencies. Moreover, efficient cooling of bulk crystals in a vacuum cryostat is achieved by adhering the back crystal surface onto a cold finger, which precludes transmission based experiments. ii) Obliquely incident and reflected light must be used in order to have sensitivity to all tensor components. This requires the crystal surface normal to be aligned exactly parallel to the rotation axis so as to maintain a constant angle of incidence. This is important because the nonlinear optical conversion efficiency is sensitive to the angle of incidence, and because the reflected beam should not precess with $\phi$ in order for it to remain stationary on the photo detector active area, which can often have a position dependent sensitivity. iii) Typical bulk single crystals of correlated electron materials may be very small ($<$ 1mm), spatially inhomogeneous and multi-faceted. To probe a small, locally flat and clean region of the crystal, that region must be made to lie exactly on the rotation axis and be coincident with the beam focus in order to avoid beam walking away from the region. That region must also be oriented normal to the rotation axis for reasons already discussed. In addition to being an alignment challenge, this would also require a cryostat manipulator with many mechanical degrees of freedom, which greatly limits the base temperature that can be reached. iv) For low temperature experiments that require low optical fluence, high harmonic signals need to be enhanced using pulsed lasers with tunable wavelength to exploit resonance conditions (eqn.~\ref{eq:microchi2}). v) Experiments that require directing an external magnetic or strain field along a particular crystallographic direction are complicated by the need to rotate the field together with the crystal, which require expensive vector magnets or rotatable strain apparatus.

\begin{figure*}
\includegraphics[scale=0.8]{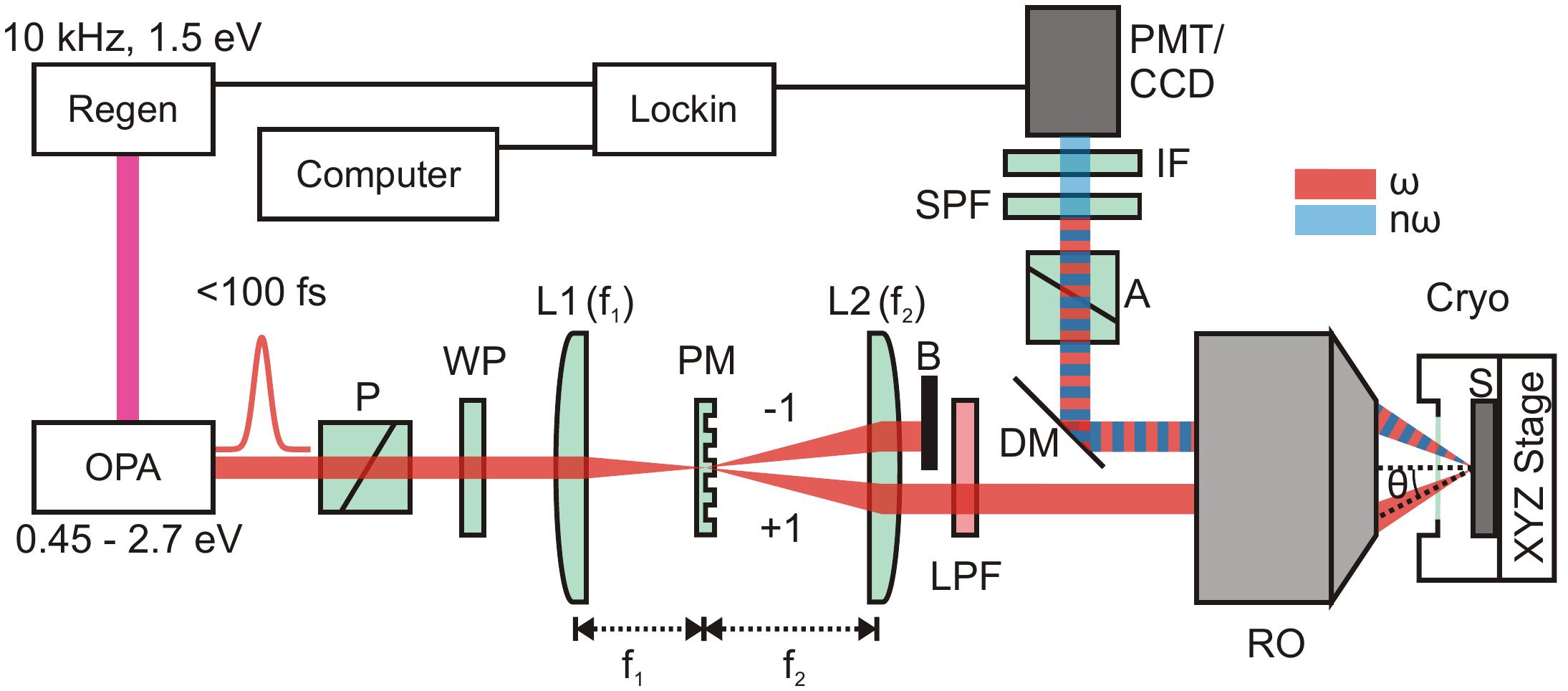}
\caption{\label{fig:2d} Layout of the NHG-RA experiment. A pulsed laser beam from an OPA, which is seeded by a Ti:sapph regenerative amplifier, passes through a polarizer (P) and waveplate (WP) and is focused by the first lens (L1) onto a phase mask (PM). A 1$^{st}$ order diffracted beam is collimated by a second lens (L2), sent through a long-pass filter (LPF), and then focused using a reflective objective (RO) onto a sample (S) in the cryostat that is mounted on an XYZ tip-tilt stage. The reflected beam passes back through the RO and is picked off by a d-cut mirror (DM). An analyzer (A), short-pass filters (SPF) and interference filter (IF) select a polarization component of the $n^{th}$ harmonic, which is measured with a photomultiplier tube (PMT) using lock-in detection or with a CCD camera. The scattering plane is rotated by placing the optics WP, PM, DM, A, SPF, IF and PMT on rotation stages. To collect the images shown in Fig. 4, the beam block (B) and DM were removed and a pellicle beamsplitter was inserted between the LPF and RO to reflect both diffracted orders through a converging lens onto a CCD camera.}
\end{figure*}

\section{Experimental System Design}\label{sec:design}

Here we describe a novel design for performing wavelength tunable NHG-RA measurements under oblique incidence geometry that meets the aforementioned technical requirements. Our scheme works by rotating the light scattering plane while keeping the crystal stationary and demonstrates both negligible beam walk on the crystal ($\leq1~\mu$m) and negligible deviation ($\leq0.2^{\circ}$) of the crystal surface normal away from the rotation axis over the entire 360$^{\circ}$ angular $\phi$ range. This opens the possibility of applying NHG-RA to small bulk single crystals and the study of their crystallographic and electronic symmetries and domain structures at ultra low temperatures, high magnetic fields and strain fields.

We used a regeneratively amplified Ti:sapphire laser system (KM Labs Wyvern-1000) producing 35~fs duration 1~mJ pulses centered at 800~nm and operating at a 10~kHz repetition rate. A pulsed laser source is exploited for its high peak fields owing to the typically low NHG efficiency of materials. For the 800~nm/400~nm second harmonic generation experiments, less than 1~mW (100~nJ/pulse) average incident power was used in order to avoid photoinduced sample damage. When operating at other wavelengths,
the full laser power pumped an optical parametric amplifier (OPA; Continuum Laser Palitra), allowing access to wavelengths ranging from approximately 500~nm to $3~\mu$m with pulse duration roughly matching that of the driving laser field. The output beam was attenuated by reflection from a $10^{\circ}$ wedge prism, which provided a large $(\sim97\%)$ ghost free reduction of the laser power without introducing unwanted pulse-broadening effects. The beam was then further attenuated by reflective neutral density filters to avoid sample damage and then delivered to the apparatus described below.

\begin{figure}
\includegraphics[scale=1.2]{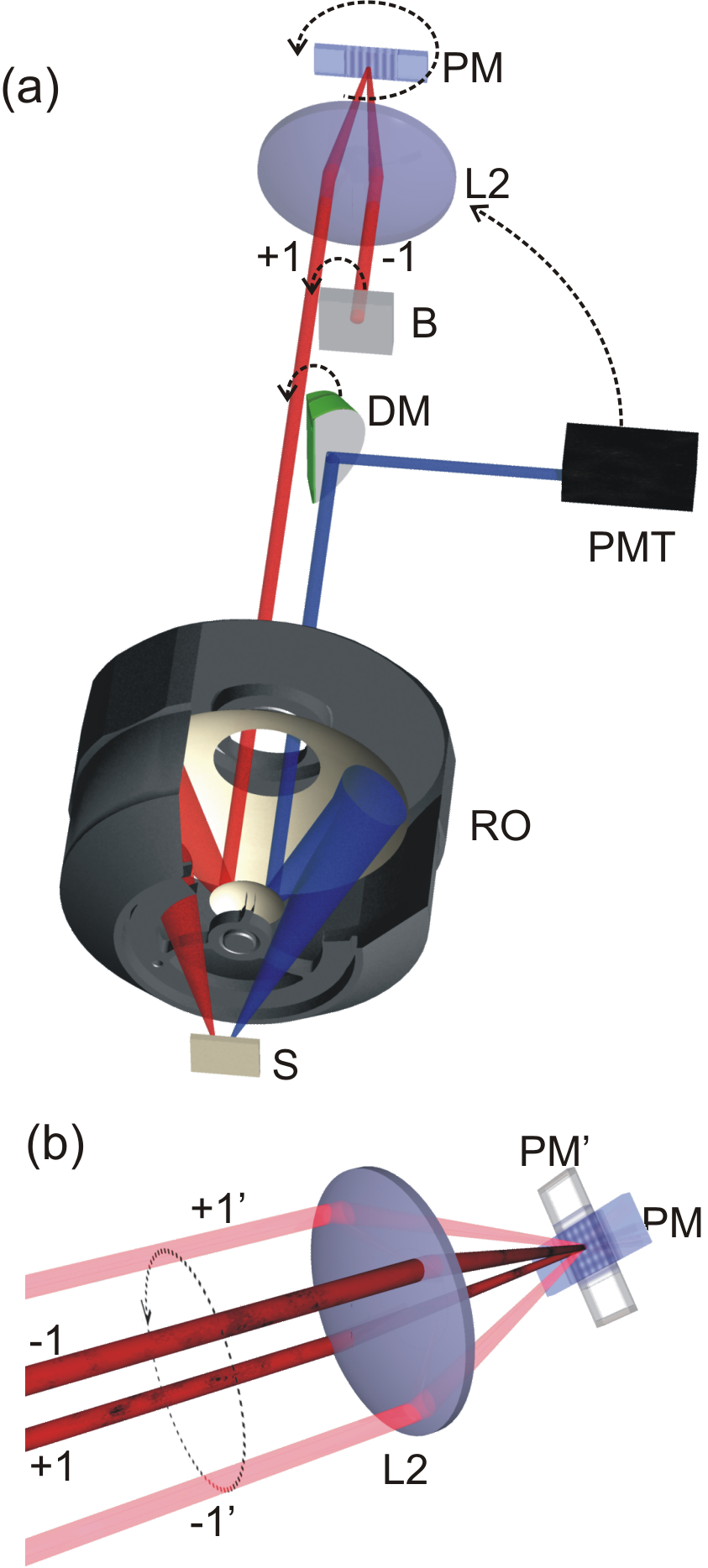}
\caption{\label{fig:setup} (a) Perspective schematic depiction of the NHG-RA system. Shown here are the diffractive binary phase mask (PM), beam block (B), collimating lens (L2), Cassegrain reflective objective (RO), sample (S), d-cut mirror (DM) and photomultiplier tube (PMT). The incident and reflected beams are shown in red and blue respectively. The dashed arrows indicate the rotating optics, with all axes of rotation coincident with the optical axis. We note that the waveplate (not shown) must also rotate to set the appropriate polarization. (b) Perspective view of only the PM and L2 showing two different rotation angles of the PM. In position PM (PM'), +1(+1') and -1(-1') orders are diffracted and collimated.}
\end{figure}

Figures~\ref{fig:2d} and \ref{fig:setup} depict our NHG-RA system, which is similar to the 4$f$ optical setups used for transient grating spectroscopy~\cite{Torchinsky}. The beam first passes through a Glan Taylor or nanoparticle polarizer (P) and then an achromatic half-waveplate (WP). It is then focused by a plano-convex lens (L1) onto a custom fused silica binary phase mask (PM - Tessera), which diffracts it equally into +1 and -1 orders at an angle $\psi$ relative to the optical axis given by $\Lambda_{PM}=\lambda/2\sin(\psi)$ where $\lambda$ is the incident wavelength and $\Lambda_{PM}$ is the feature size on the PM. An array of feature sizes adapted for different incident wavelength ranges are available on our PM. Both diffracted orders are simultaneously collimated and brought parallel to each other
by an achromatic doublet (L2), which was chosen for both its reduced chromatic and optical abberations over the wavelength range of the incident light. One order is then blocked by a beam block (B) while the other passes through a longpass filter (LPF) to block parasitic higher harmonics. The final optical element in the light incidence path is a 15$\times$, infinite back focal length Cassegrain reflective objective (RO) with a UV-enhanced Al coating that serves to focus the light onto the sample without chromatic dispersion, spherical abberation, coma and astigmatism, significantly loosening the alignment tolerances of this component of the experiment. This optic also provides a large numerical aperture (NA=0.5), yielding an oblique incidence angle of $\theta\sim~30^\circ$ onto the sample at a working distance of 25~mm, which exceeds the minimum working distance of our optical vacuum cryostat (Janis ST-500). The optical cryostat is mounted on a custom stage with XYZ translational and tip-tilt angular degrees of freedom for sample alignment.

The fundamental and higher harmonic beams reflected from the sample all follow an equal path back through the RO that is diametrically opposite from the incident beam since the RO is free of chromatic dispersion and abberation for all wavelengths used. A d-cut silver coated pick-off mirror (DM) steers the reflected beams through a high contrast ratio analyzer (A) to select either the P or S output polarization (Fig.~\ref{fig:splane}a), which is then spectrally filtered using two consecutive shortpass filters (SPF) and an interference filter (IF) at the desired harmonic frequency and finally directed into a photomultiplier tube (PMT). We note that a beam diffuser can be placed just before the PMT to more uniformly illuminate the PMT active area. The intensity is detected by terminating the output current of the PMT across a 50~k$\Omega$ resistor and inputting to a lock-in amplifier synchronized with the repetition rate of the laser.

In order to rotate the scattering plane, a subset of the optics are placed on motorized rotation stages that share a common axis of rotation along the optical axis (stages not shown in Figs ~\ref{fig:2d} and \ref{fig:setup}). Specifically, the WP is mounted on a dedicated rotation stage to maintain either P or S polarized incident light with respect to the rotating scattering plane; the PM is mounted on a second rotation stage such that the diffracted beams draw a cone under rotation (Fig.~\ref{fig:setup}b); and the DM, A, SPF, IF and PMT are mounted together on a third rotation stage to track the displacement of the reflected beam. RA patterns are collected by stepping the
rotation angles in finite increments (the WP stage is stepped at half increments of the other two stages) and taking measurements at each angle. Typical collection times for the data shown in Section V were on the order of 45 minutes for a full $\phi$-dependent trace.

\section{System Performance}\label{sec:results}

\subsection{Performance parameters}\label{sec:param}

The optics L2 and RO comprise the two elements of a Keplerian telescope, which serves to image the laser spot on the PM onto the surface of the sample. When B is removed so that the +1 and -1 diffracted orders are allowed to recombine at the surface of the sample, the phase object of the binary mask pattern is converted into an amplitude image in the form of a sinusoidal interference pattern, whose fringe spacing $\Lambda$ is related to the angle of incidence by $\Lambda=\lambda/2\sin(\theta)$. As the scattering plane and orientation of the interference fringes rotate with the PM, the amount of beam walk on the sample and any variation in the scattering angle over the 360$^{\circ}$ angular $\phi$ range can be quantified by tracking the location of the interference pattern and the magnitude of $\Lambda$ respectively.

To perform these tests, we removed B and DM and placed a pellicle beam splitter in between L2 and the RO. After being collimated by L2, both +1 and -1 diffracted orders pass through the pellicle into the RO and then converge at their focus on the sample surface. Both +1 and -1 beams then reflect off of the sample, are re-collimated through the RO, and are steered by the pellicle into an achromatic doublet that focuses them onto a CCD camera. To ensure that the area on the sample illuminated by the laser beams is oriented normal to the optical rotation axis, the reflected +1(-1) beam path is made to completely overlap the incident -1(+1) order beam path for all $\phi$. We verify that both +1 and -1 orders independently provide the same sharp image of the sample surface and that they overlap entirely with each other on the CCD camera. Using 800~nm incident light and a phase mask feature size of $\Lambda_{PM}$ = 13.4~$\mu$m, we obtain $\Lambda \sim 900$~nm and an overall spot size on the sample less than $20~\mu$m at FWHM. We note that it is possible to achieve smaller spot sizes on the sample simply by decreasing the focal length of L1 to shrink the laser spot size on the PM. However, the effects of an increasingly large longitudinal field component of a focused vector Gaussian field~\cite{Quesnel, Carrasco} should be considered when analyzing the NHG patterns.

\begin{figure}
\includegraphics[scale=0.68]{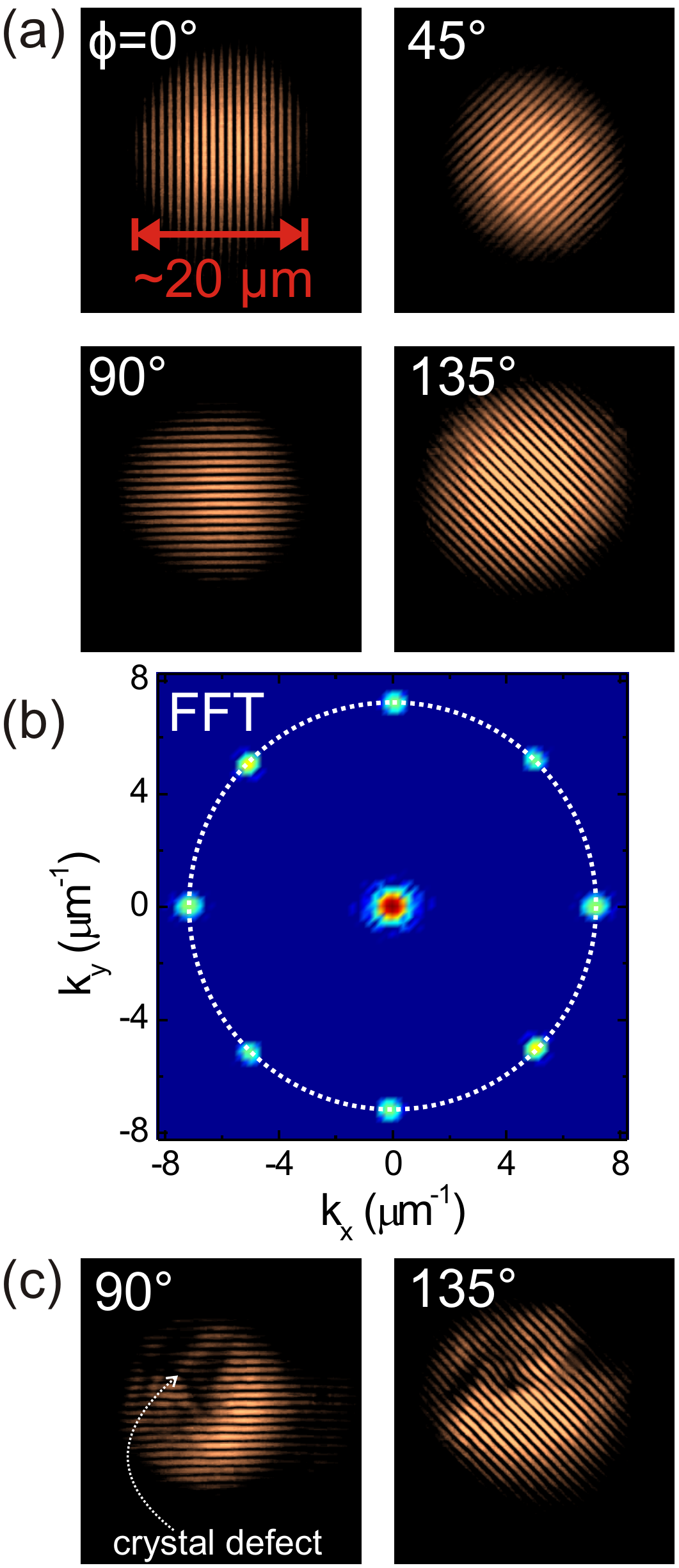}
\caption{\label{fig:rotate}(a) Images generated through the interference of both the +1 and -1 orders ($\lambda$ = 800~nm) on the surface of a sample as the phase mask is rotated. Both the interference fringe spacing ($\sim0.9~\mu$m) and the overall beam diameter ($\sim20~\mu$m) show negligible changes with $\phi$. (b) Superposition of the spatial Fourier transforms of images such as those shown in panel (a). The magnitudes of the fringe wave vectors all lie on a circle (white dotted line). (c) Interference image taken near a surface defect. The defect location relative to the beam spot is unaltered to within the precision of the CCD image, setting an upper-bound of the beam walk on the sample surface to be $\sim 1~\mu$m. Note that the slight changes to the appearance of the defect between $\phi = 90^{\circ}$ and $\phi = 135^{\circ}$ is due to its 3D nature, which causes it to cast different shadows when it is illuminated by the laser from different angles.}
\end{figure}

Interference patterns on the sample surface at various values of $\phi$ are shown in Fig.~\ref{fig:rotate}a, which show no appreciable change in $\Lambda$. This is more clearly demonstrated by taking their Fourier transforms (Fig.~\ref{fig:rotate}b), which show a central peak at $\vec{k}=0$ due to the FFT of the overall beam shape and satellite peaks at $|k| = 7~\mathrm{\mu m^{-1}}$ representing the modulation wave vector of the interference fringes. The position of the satellite peaks is carefully tracked as the scattering plane is rotated. Using the derived values of $\Lambda$, we determine its change with $\phi$ to be less than $\sim 0.01~\mu$m, which corresponds to a variation in scattering angle $\delta\theta < \pm0.2^{\circ}$.

To determine the amount of beam walking on the sample as the scattering plane is rotated, we use the presence of defects on a sample surface to serve as a point of reference. In general we find that the location of defects are stationary relative to the edges of the interference pattern to within $1~\mu$m as $\phi$ is varied. An example of a large defect is shown in Fig.~\ref{fig:rotate}c. Taking the metrics of Figs.~\ref{fig:rotate}b and c together, our NHG-RA setup features an incident beam that stays highly stationary on the sample and exhibits negligible variation in its incidence angle on the sample over the full range of $\phi$.

One experimental inconvenience of our scheme is that the diverging reflector in the RO is suspended by a ``spider" mount which occludes the beam in four angular positions separated by $90^\circ$. The angular subtense of this occlusion is $\pm 8^\circ$ in our current configuration but can be further reduced by decreasing the collimated laser beam diameter emerging from L2. To eliminate the occluded angles, we chose to mount the RO in a precision manual rotation stage. Each RA pattern is taken twice with the RO rotated to two different angles and then patched together as discussed in Sec.~\ref{sec:example}. Alternatively, the RO can be mounted on a motorized rotation stage and simply rotated in step with the PM during data acquisition.

\subsection{Typical example of measurement on Sr$_2$IrO$_4$}\label{sec:example}

\begin{figure}
\includegraphics[scale=0.75]{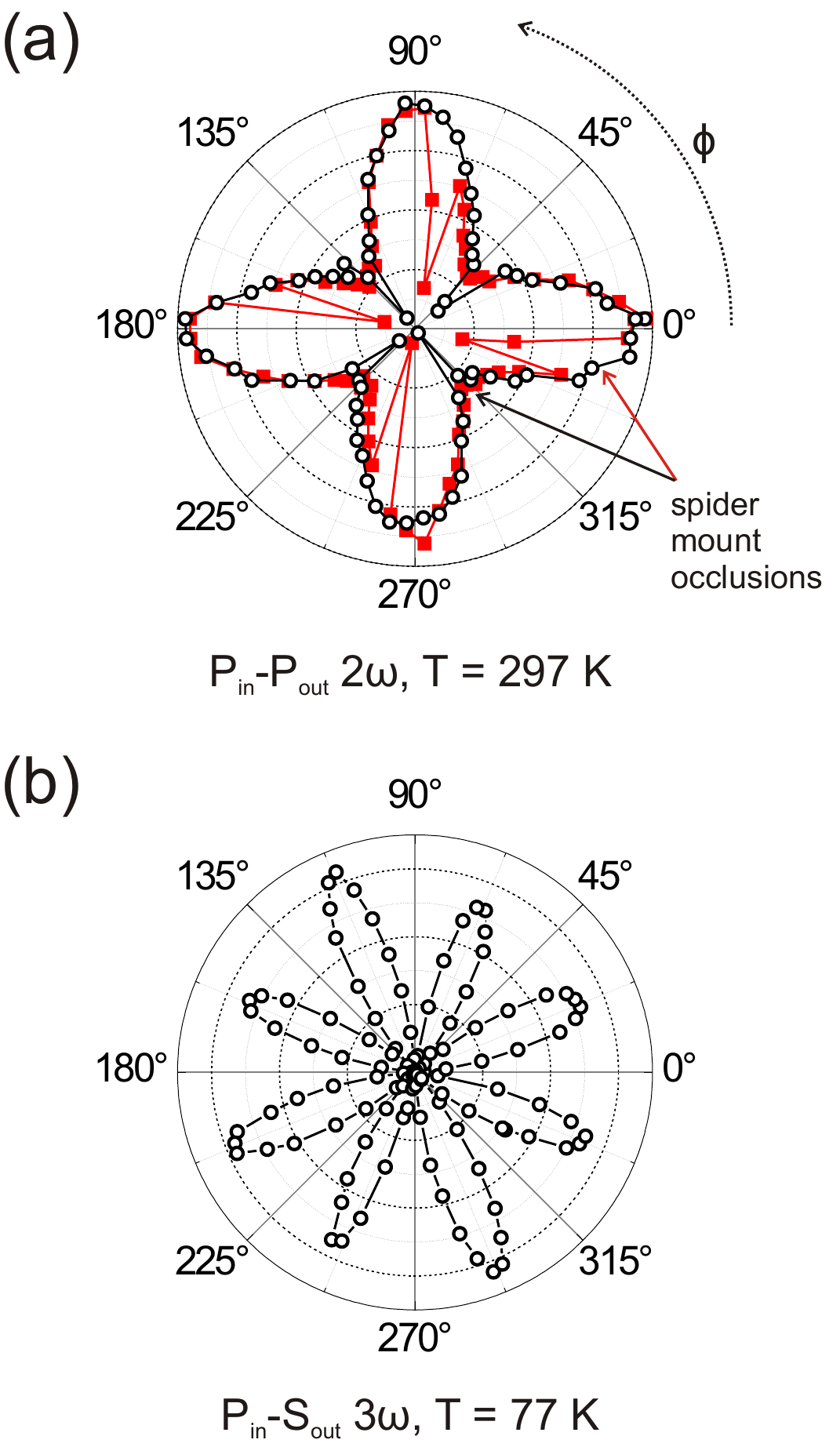}
\caption{\label{fig:flowers} Representative NHG-RA measurements on a Sr$_2$IrO$_4$ bulk single crystal. (a) 800~nm/400~nm SHG patterns taken at room temperature in PP geometry using two different orientations of the RO. (b) 1200~nm/400~nm THG pattern taken at T=77~K under PS geometry. The RO is mounted so that the spider mount occlusion coincides with nodes in the THG pattern.}
\end{figure}

To demonstrate the power of our technique, we apply our NHG-RA setup to study a single crystal of the 5$d$ transition metal oxide Sr$_2$IrO$_4$. The single crystal growth methods are described elsewhere \cite{Gang}. Iridates are generally difficult to study using neutron diffraction because Ir is a strong neutron absorber and bulk single crystals are typically small ($\leq1$~mm) and may have micron scale domains \cite{Ye, Dhital, Boseggia}. Moreover, their cleaved surfaces are often optically flat over micron scale facets and terraces. Using our setup, we are able to collect NHG-RA data from an (001) facet of a Sr$_2$IrO$_4$ single crystal under varying polarization combinations, temperatures and wavelengths as shown in Figure~\ref{fig:flowers}. Panel (a) displays room temperature second harmonic generation (SHG) 800~nm/400~nm data taken with P-polarized incident and reflected light. The signal, which originates from a bulk electric quadrupolar response $\chi^{qee}_{ijkl}$, exhibits four-fold rotational symmetry about the (001) axis as required by the Sr$_2$IrO$_4$ crystalline lattice \cite{Crawford, Huang}. Two superposed raw data sets taken at two different RO orientations are presented showing excellent reproducibility. The presence of the spider mount occlusions mentioned in Sec.~\ref{sec:param} are clearly visible as sharp valleys in the RA patterns, which can be eliminated by patching together the two data sets shown.

As an example of experiments conducted under cryogenic conditions using a tunable light source, we perform third harmonic generation (THG) 1200~nm/400~nm experiments on Sr$_2$IrO$_4$ at 77 K. NHG-RA data taken with S-polarized incident and P-polarized reflected light are shown in Fig.~\ref{fig:flowers}b. This response originates from a bulk electric dipole process $\chi^{eeee}_{ijkl}$ and the symmetry of the underlying crystalline lattice is again observed with excellent signal-to-noise contrast. In this case, the spider mount occlusions are eliminated by orienting them to coincide with the nodes in the THG pattern. The fact that all nodes approach zero indicates a negligible background noise in our data.

\section{Conclusions \& Outlook}\label{sec:summary}

The NHG-RA spectrometer developed here resolves previous technical challenges associated with beam walking on the sample and precession of the sample normal with respect to the sample rotation axis. This opens the possibility of applying nonlinear optics as a probe of lattice and electronic symmetries on small bulk single crystals in ultra low temperature, high magnetic field or high pressure environments, which can greatly complement diffraction based techniques. In particular, this method should find wide application in the characterization of temperature, magnetic field or pressure driven complex electronic phases in strongly correlated $d$- and $f$-electron systems.

The techniques used to produce the images in Fig.~\ref{fig:rotate} may be refined to perform an NHG-RA experiment in microscopy mode that combines the demonstrated advantages of our approach with diffraction-limited spatial resolution. This can facilitate the search for lattice, magnetic and even more exotic electronically ordered domains in a crystal. Given that our technique uses ultrashort laser pulses, it can also be implemented as a time-resolved pump-probe experiment by introducing a pump beam with a variable time delay. This will allow real-time observation of lattice or electronic symmetry changes following photo-excitation that can lead to a more detailed understanding of the coupling between lattice and electronic degrees of freedom in a crystal. There is also the possibility of directly observing the symmetry of coherently generated collective modes by measuring the time-resolved symmetry variations in the RA patterns following photo-excitation. These may include normal modes of lattice, magnetic or other electronic orders, such as spin and charge density waves.

\begin{acknowledgments}
D.H. acknowledges partial support by the U. S. Army Research Office under grant number W911NF-13-1-0059. Instrumentation for the NHG-RA setup was partially supported by a U. S. Army Research Office DURIP award under grant number W911NF-13-1-0293. D.H. acknowledges funding provided by the Institute for Quantum Information and Matter, an NSF Physics Frontiers Center (PHY-1125565) with support of the Gordon and Betty Moore Foundation through Grant GBMF1250. G.C. acknowledges NSF support via grant DMR-1265162.
\end{acknowledgments}


\end{document}